# Third order perturbed Heisenberg Hamiltonian of sc ferromagnetic films with fifty spin layers

N.U.S. Yapa[2, 3] and P. Samarasekara[1, 3*]

[1]Department of Physics, University of Peradeniya, Peradeniya, Sri Lanka

[2]Department of Physics, Open University of Sri Lanka, Kandy, Sri Lanka

[3]Postgraduate Institute of science, University of Peradeniya, Sri Lanka

*Abstract*

*Magnetic energy of simple cubic structured ferromagnetic films with 10 to 50 spin layers was determined using third order perturbed Heisenberg Hamiltonian. By plotting 3-D plot of energy versus angle and stress induced anisotropy, the values of stress induced anisotropy corresponding to energy minimums and maximums were determined. By plotting the graphs of energy versus angle at these different stress induced anisotropy values, the easy and hard directions were determined. Magnetic easy and hard directions are related to energy minimums and maximums of the curve. Similar graphs were plotted for spin exchange interaction to determine the easy and hard directions. Graphs of energy versus angle were plotted by keeping all the magnetic energy parameters at constant values for each number of spin layers to determine the variation of magnetic easy directions, hard direction and corresponding energies with the number of spin layers. The magnetic easy axis gradually rotates from out of plane to in plane direction as the number of spin layers is increased. In addition, the magnetic anisotropy energy increases with the number of spin layers.*

## 1. Introduction:

Second and third order perturbed Heisenberg Hamiltonian has been employed to find the magnetic properties of ferromagnetic and ferrite films. The interfacial coupling dependence of the magnetic ordering in ferro-antiferromagntic bilayers has been studied using the Heisenberg Hamiltonian [1]. Heisenberg Hamiltonian incorporated with spin exchange interaction, magnetic dipole interaction, applied magnetic field, second and fourth order magnhetic anisotropy terms has been solved for ferromagnetic thin films [2, 3, 4]. The domain structure and Magnetization reversal in thin magnetic films was described using computer simulations [5]. Heisenberg

Hamiltonian was used to describe in-plane dipole coupling anisotropy of a square ferromagnetic Heisenberg monolayer [6]. In addition, some other models can be summarized as following. The quasistatic magnetic hysteresis of ferromagnetic thin films grown on a vicinal substrate has been theoretically investigated using Monte Carlo simulations [7]. Structural and magnetic properties of two dimensional FeCo ordered alloys have been determined by first principles band structure theory [8]. EuTe films with surface elastic stresses have been theoretically studied using Heisenberg Hamiltonian [9]. De Vries theory was employed to explain the magnetostriction of dc magnetron sputtered FeTaN thin films [10]. Magnetic layers of Ni on Cu have been theoretically investigated using the Korringa-Kohn-Rostoker Green's function method [11]. Electric and magnetic properties of multiferroic thin films have been theoretically explained using modified Heisenberg model and transverse Ising model coupled with Green's function technique [12].

Our previous research work can be summarized as following. Second order perturbed Heisenberg Hamiltonian of ferromagnetic films with four spin layers was determined under special assumptions to avoid tedious derivations [13]. Second order perturbed Heisenberg Hamiltonian of ferromagnetic films of five spin layers with all seven magnetic energy parameters was solved without any special assumptions [14]. The third order perturbed Heisenberg Hamiltonian was solved for thick ferromagnetic films under several special assumptions [15]. The third order perturbed Heisenberg Hamiltonian with all seven magnetic energy parameters was solved for ferromagnetic films with three spin layers [16]. In addition, the third order perturbed Heisenberg Hamiltonian was also used to explain the magnetic properties of ferrites [17]. Furthermore, the magnetic dipole interaction was calculated for ferromagnetic cobalt with a complicated structure [18].

In this manuscript, $3^{rd}$ order perturbed Heisenberg Hamiltonian with all seven magnetic energy parameters was solved for simple cubic (sc) structured ferromagnetic films with spin layers of 10 to 50. Variation of magnetic easy axis and the total magnetic energy with number of layers was investigated. MATLAB computer program was employed to plot the 2-D and 3-D graphs for different values of magnetic energy parameters.

## 2. Model:

The Heisenberg Hamiltonian of ferromagnetic films can be written as following.

$$H = -\frac{J}{2}\sum_{m,n}\vec{S}_m.\vec{S}_n + \frac{\omega}{2}\sum_{m\ne n}(\frac{\vec{S}_m.\vec{S}_n}{r_{mn}^3} - \frac{3(\vec{S}_m.\vec{r}_{mn})(\vec{r}_{mn}.\vec{S}_n)}{r_{mn}^5}) - \sum_m D_{\lambda_m}^{(2)}(S_m^z)^2 - \sum_m D_{\lambda_m}^{(4)}(S_m^z)^4$$

$$-\sum_{m,n}[\vec{H} - (N_d\vec{S}_n/\mu_0)].\vec{S}_m - \sum_m K_s Sin2\theta_m$$

After using spins with unit magnitudes, the equation will be deduced to following form.

$$E(\theta) = -\frac{1}{2}\sum_{m,n=1}^{N}[(JZ_{|m-n|} - \frac{\omega}{4}\Phi_{|m-n|})\cos(\theta_m - \theta_n) - \frac{3\omega}{4}\Phi_{|m-n|}\cos(\theta_m + \theta_n)]$$

$$-\sum_{m=1}^{N}(D_m^{(2)}\cos^2\theta_m + D_m^{(4)}\cos^4\theta_m + H_{in}\sin\theta_m + H_{out}\cos\theta_m)$$

$$+\sum_{m,n=1}^{N}\frac{N_d}{\mu_0}\cos(\theta_m - \theta_n) - K_s\sum_{m=1}^{N}\sin 2\theta_m \quad (1)$$

Here N, m (or n), J, $Z_{|m-n|}$, $\omega$, $\Phi_{|m-n|}$, $\theta_m(\theta_n)$, $D_m^{(2)}$, $D_m^{(4)}$, $H_{in}$, $H_{out}$, $N_d$ and $K_s$ are total number of layers, layer index, spin exchange interaction, number of nearest spin neighbors, strength of long range dipole interaction, partial summations of dipole interactions, azimuthal angles of spins, second and fourth order anisotropy constants, in plane and out of plane applied magnetic fields, demagnetization factor and stress induced anisotropy constants, respectively.

By choosing azimuthal angles as $\theta_m = \theta + \varepsilon_m$ and $\theta_n = \theta + \varepsilon_n$, above energy can be expanded up to the third order of $\varepsilon$ as following,

$$E(\theta) = E_0 + E(\varepsilon) + E(\varepsilon^2) + E(\varepsilon^3) \quad (2)$$

Here $E_0 = -\frac{1}{2}\sum_{m,n=1}^{N}(JZ_{|m-n|} - \frac{\omega}{4}\Phi_{|m-n|}) + \frac{3\omega}{8}\cos 2\theta \sum_{m,n=1}^{N}\Phi_{|m-n|}$

$$-\cos^2\theta\sum_{m=1}^{N}D_m^{(2)} - \cos^4\theta\sum_{m=1}^{N}D_m^{(4)} - N(H_{in}\sin\theta + H_{out}\cos\theta - \frac{N_d}{\mu_0} + K_s\sin 2\theta) \quad (3)$$

$$E(\varepsilon) = -\frac{3\omega}{8}\sin 2\theta\sum_{m,n=1}^{N}\Phi_{|m-n|}(\varepsilon_m + \varepsilon_n) + \sin 2\theta\sum_{m=1}^{N}D_m^{(2)}\varepsilon_m + 2\cos^2\theta\sin 2\theta\sum_{m=1}^{N}D_m^{(4)}\varepsilon_m$$

$$-H_{in}\cos\theta\sum_{m=1}^{N}\varepsilon_m + H_{out}\sin\theta\sum_{m=1}^{N}\varepsilon_m - 2K_s\cos 2\theta\sum_{m=1}^{N}\varepsilon_m$$

$$E(\varepsilon^2) = \frac{1}{4} \sum_{m,n=1}^{N} (JZ_{|m-n|} - \frac{\omega}{4} \Phi_{|m-n|})(\varepsilon_m - \varepsilon_n)^2 - \frac{3\omega}{16} \cos 2\theta \sum_{m,n=1}^{N} \Phi_{|m-n|} (\varepsilon_m + \varepsilon_n)^2$$

$$- (\sin^2 \theta - \cos^2 \theta) \sum_{m=1}^{N} D_m^{(2)} \varepsilon_m^2 + 2\cos^2 \theta (\cos^2 \theta - 3\sin^2 \theta) \sum_{m=1}^{N} D_m^{(4)} \varepsilon_m^2$$

$$+ \frac{H_{in}}{2} \sin \theta \sum_{m=1}^{N} \varepsilon_m^2 + \frac{H_{out}}{2} \cos \theta \sum_{m=1}^{N} \varepsilon_m^2 - \frac{N_d}{2\mu_0} \sum_{m,n=1}^{N} (\varepsilon_m - \varepsilon_n)^2$$

$$+ 2K_s \sin 2\theta \sum_{m=1}^{N} \varepsilon_m^2$$

$$E(\varepsilon^3) = \frac{\omega}{16} \sin 2\theta \sum_{m,n=1}^{N} (\varepsilon_m + \varepsilon_n)^3 \phi_{|m-n|} - \frac{4}{3} \cos \theta \sin \theta \sum_{m=1}^{N} D_m^{(2)} \varepsilon_m^3$$

$$- 4\cos \theta \sin \theta (\frac{5}{3} \cos^2 \theta - \sin^2 \theta) \sum_{m=1}^{N} D_m^{(4)} \varepsilon_m^3 + \frac{H_{in}}{6} \cos \theta \sum_{m=1}^{N} \varepsilon_m^3$$

$$- \frac{H_{out}}{6} \sin \theta \sum_{m=1}^{N} \varepsilon_m^3 + \frac{4K_s}{3} \cos 2\theta \sum_{m=1}^{N} \varepsilon_m^3$$

After using the constraint $\sum_{m=1}^{N} \varepsilon_m = 0$, $E(\varepsilon) = \vec{\alpha}.\vec{\varepsilon}$

Here $\vec{\alpha}(\varepsilon) = \vec{B}(\theta) \sin 2\theta$ are the terms of matrices with

$$B_\lambda(\theta) = -\frac{3\omega}{4} \sum_{m=1}^{N} \Phi_{|\lambda-m|} + D_\lambda^{(2)} + 2D_\lambda^{(4)} \cos^2 \theta \qquad (4)$$

Also $E(\varepsilon^2) = \frac{1}{2} \vec{\varepsilon}.C.\vec{\varepsilon}$

Here the elements of matrix C can be given as following,

$$C_{mn} = -(JZ_{|m-n|} - \frac{\omega}{4} \Phi_{|m-n|}) - \frac{3\omega}{4} \cos 2\theta \Phi_{|m-n|} + \frac{2N_d}{\mu_0}$$

$$+ \delta_{mn} \{ \sum_{\lambda=1}^{N} [JZ_{|m-\lambda|} - \Phi_{|m-\lambda|}(\frac{\omega}{4} + \frac{3\omega}{4} \cos 2\theta)] - 2(\sin^2 \theta - \cos^2 \theta) D_m^{(2)}$$

$$+ 4\cos^2 \theta (\cos^2 \theta - 3\sin^2 \theta) D_m^{(4)} + H_{in} \sin \theta + H_{out} \cos \theta - \frac{4N_d}{\mu_0} + 4K_s \sin 2\theta \} \quad (5)$$

If $E(\varepsilon^3) = \varepsilon^2 \beta.\vec{\varepsilon}$, then matrix elements of matrix β can be given as following,

$$\beta_{mn} = \frac{3\omega}{8}\sin 2\theta \Phi_{|m-n|} + \delta_{mn}\{\frac{\omega}{8}\sin 2\theta [A_m - \Phi_0] - \frac{4}{3}\cos\theta\sin\theta D_m^{(2)}$$

$$- 4\cos\theta\sin\theta(\frac{5}{3}\cos^2\theta - \sin^2\theta)D_m^{(4)} + \frac{H_{in}}{6}\cos\theta - \frac{H_{out}}{6}\sin\theta$$

$$+ \frac{4K_s}{3}\cos 2\theta\} \qquad (6)$$

Also $\beta_{nm}=\beta_{mn}$ and matrix $\beta$ is symmetric.

Here $A_m$ values are different for even and odd N values, and can be given as following.

For odd N, $A_{\frac{N}{2}+0.5+n} = 2\sum_{\nu=0}^{\frac{N}{2}-0.5-n}\Phi_\nu + \sum_{\nu=\frac{N}{2}+0.5-n}^{\frac{N}{2}+n-0.5}\Phi_\nu$ for $m > \frac{N}{2}$

Here n varies from 1 to $\frac{N}{2}-0.5$.

When n=0, $A_{\frac{N}{2}+0.5+n} = 2\sum_{\nu=0}^{\frac{N}{2}-0.5-n}\Phi_\nu$

$A_m$ for $m < \frac{N}{2}$ can be found using $A_{\frac{N}{2}+0.5+n} = A_{\frac{N}{2}+0.5-n}$

For even N, $A_{\frac{N}{2}+1+n} = 2\sum_{\nu=0}^{\frac{N}{2}-1-n}\Phi_\nu + \sum_{\nu=\frac{N}{2}-n}^{\frac{N}{2}+n}\Phi_\nu$ for $m > \frac{N}{2}$ (6a)

Here n varies from 0 to $\frac{N}{2}-1$.

$A_m$ for $m < \frac{N}{2}$ can be obtained using $A_{\frac{N}{2}+1+n} = A_{\frac{N}{2}-n}$ (6b)

Therefore, the total magnetic energy given in equation 2 can be deduced to

$E(\theta)=E_0+ \vec{\alpha}.\vec{\varepsilon} + \frac{1}{2}\vec{\varepsilon}.C.\vec{\varepsilon} + \varepsilon^2 \beta.\vec{\varepsilon}$ (7)

Because it is difficult to find an equation for ε in the presence of the third order of ε in above equation, only the second order of ε will be considered for following derivation.

Then $E(\theta) = E_0 + \vec{\alpha}.\vec{\varepsilon} + \frac{1}{2}\vec{\varepsilon}.C.\vec{\varepsilon}$

Using a proper constraint provides $\vec{\varepsilon} = -C^+.\vec{\alpha}$

Here $C^+$ is the pseudo-inverse given by

$$C.C^+ = 1 - \frac{E}{N}. \qquad (8)$$

E is the matrix with all elements given by $E_{mn}=1$.

After using ε in equation 7, $E(\theta) = E_0 - \frac{1}{2}\vec{\alpha}.C^+.\vec{\alpha} - (C^+\alpha)^2 \vec{\beta}(C^+\alpha) \qquad (9)$

Above equation gives the energy per unit spin.

## 3. Result and Discussion:

After finding matrix elements from equations 4, 5 and 6, the total magnetic energy was determined using equation 9. For s. c. (001) lattice, $Z_0=4$, $Z_1=1$, $\Phi_0=9.0336$ and $\Phi_1=-0.3275$ [2, 3, 4]. Figure 1 exhibits the 3-D plot of energy versus stress induced anisotropy and angle for sc structures ferromagnetic film with 40 spin layers. Other values were fixed at $\frac{J}{\omega} = \frac{H_{in}}{\omega} = \frac{N_d}{\mu_0 \omega} = \frac{H_{out}}{\omega} = 10$, $\frac{D_m^{(2)}}{\omega} = 30$ and $\frac{D_m^{(4)}}{\omega} = 20$. Several energy minimums and maximums can be observed at different values of $\frac{K_s}{\omega}$ and angles in this 3-D plot. One maximum and one minimum of this 3-D plot could be observed at $\frac{K_s}{\omega} = 16$ and $\frac{K_s}{\omega} = 9$, respectively. When the film sample is cooled or heated before or after deposition or annealing, the macroscopic level stress is induced in the film. Due to this stress, an extra anisotropy is induced in the film, and hence the coercivity of the film changes [19, 20, 21]. The same 3-D plot of energy versus stress induced anisotropy and angle was plotted for sc lattice with four spin layers using 2[nd] order perturbed Heisenberg Hamiltonian, and the shape of that graph is entirely different from the graph given in this manuscript for N=40 [13]. In addition, the energy of the same graph for four layered case changes only up to 1000. Because the total number of spins in film increases with number of spin layers, the energy of a film with more spin layers must be simply higher.

The same graph for fcc structured ferromagnetic film with four and five spin layers was plotted using 2$^{nd}$ order perturbed Heisenberg Hamiltonian by us [14]. Those graphs for four and five spin layers are also entirely different from the graph given for N=40 in this manuscript. Energy in those graphs varies only up to 800. The shape of the same 3-D graph obtained using 3$^{rd}$ order perturbed Heisenberg Hamiltonian of fcc structured ferromagnetic films with three spin layers was different [16]. Energy in that case varies only up to $10^4$.

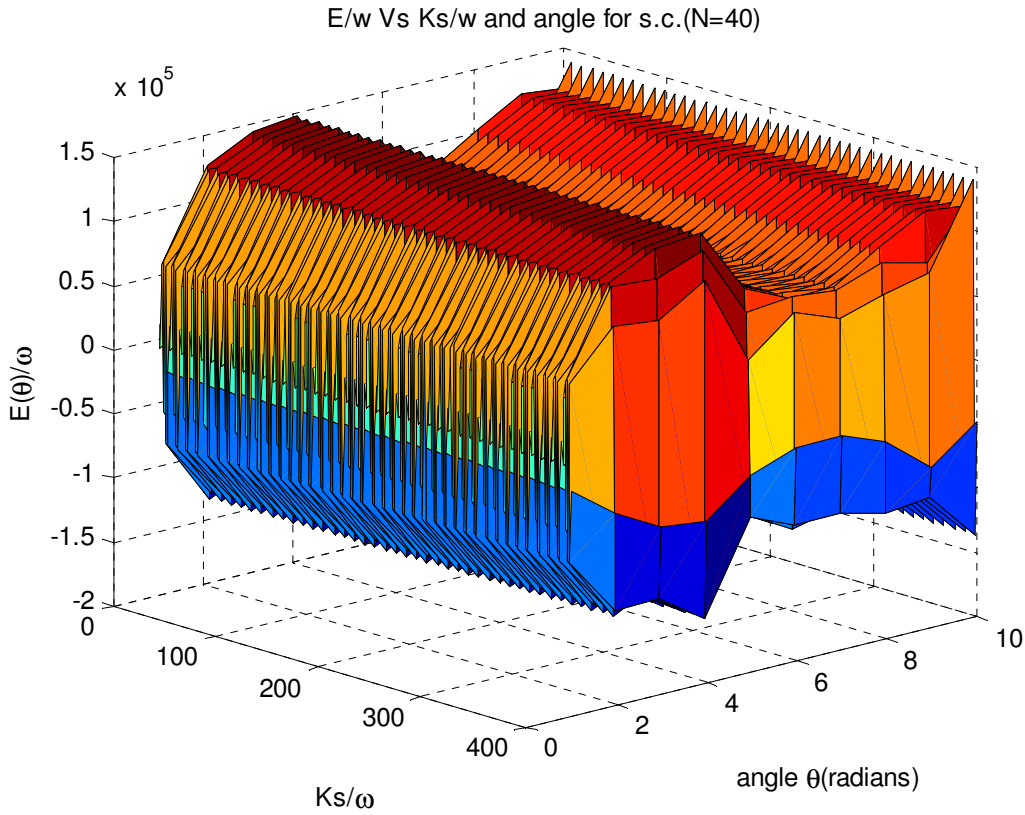

Figure 1: 3-D plot of $\dfrac{E(\theta)}{\omega}$ versus angle and $\dfrac{K_s}{\omega}$ for N=40.

Figure 2 shows the graph of energy versus angle at $\dfrac{K_s}{\omega}$=16 for N=40. A minimum and a maximum of this plot can be observed at 16.1975$^0$ and 120.6019$^0$, respectively. Magnetic hard axis can be observed at 120.6019$^0$. Figure 3 shows the graph of energy versus angle at $\dfrac{K_s}{\omega}$=9. A minimum and a maximum of this plot can be observed at 10.8002$^0$ and 115.1989$^0$, respectively. Magnetic easy axis can be observed at 10.8002$^0$.

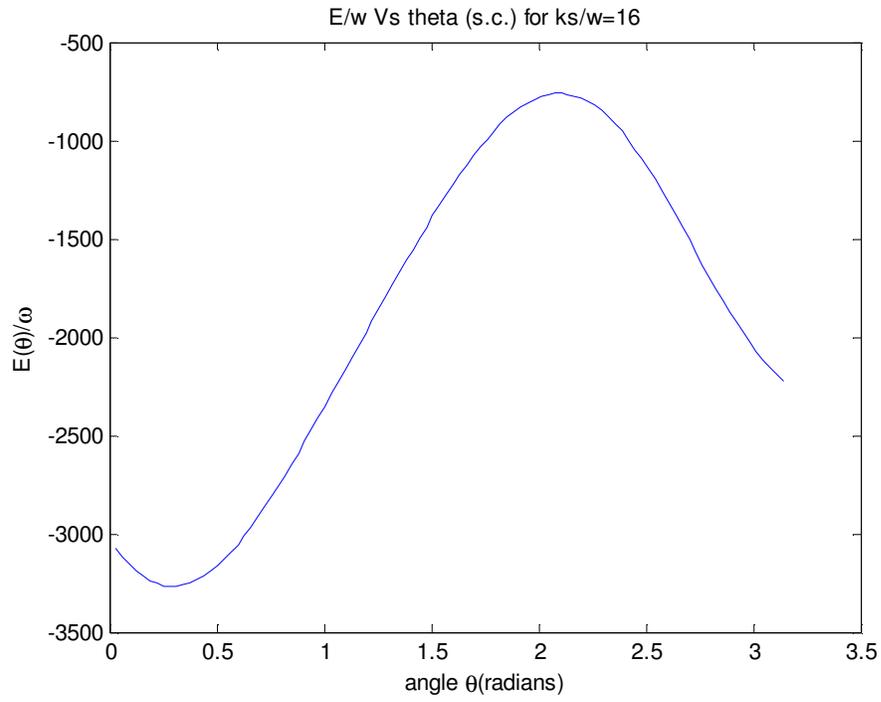

Figure 2: Energy versus angle at $\frac{K_s}{\omega}=16$.

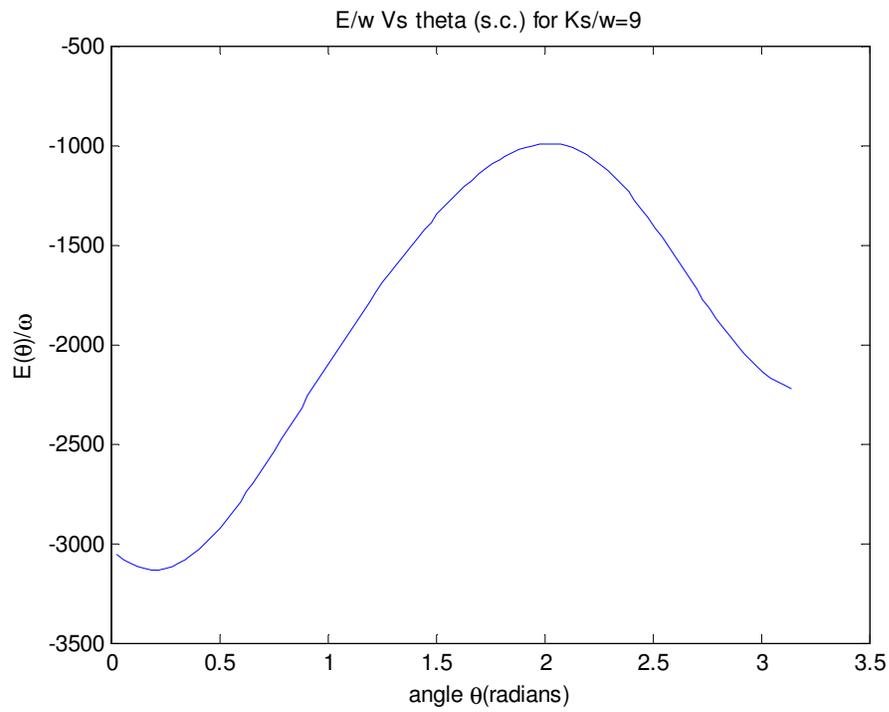

Figure 3: Energy versus angle at $\frac{K_s}{\omega}=9$.

Figure 4 is the 3-D plot of energy versus spin exchange interaction and angle. Among several maximums and minimums in this 3-D plot, one minima and maxima can be observed at $\frac{J}{\omega}=4$ and $\frac{J}{\omega}=5$, respectively. According to the graph of energy versus angle at $\frac{J}{\omega}=4$, a minimum and a maximum can be observed at $12.5993^0$ and $116.9980^0$, respectively. As a result, magnetic easy axis is along $12.5993^0$. According to the graph of energy versus angle at $\frac{J}{\omega}=5$, a minimum and a maximum of can be observed at $12.5993^0$ and $115.2218^0$, respectively. Magnetic hard axis is along $115.2218^0$. This same 3-D graph plotted for fcc structured ferromagnetic films with three spin layers using 3rd order perturbed Heisenberg Hamiltonian was different from the 3-D plot given here [16]. Although the overall energy gradually decreases with the increase of $\frac{J}{\omega}$ in this graph, the energy shows only the periodic change in the graph given in other manuscript for N=3. In addition, energy in N=3 graph varies from -1000 to 100 only.

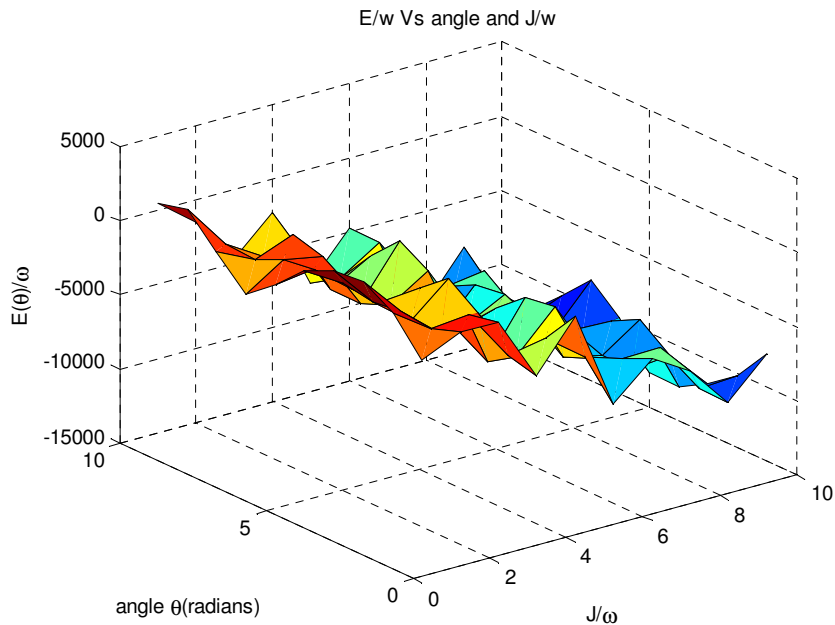

Figure 4: 3-D plot of $\frac{E(\theta)}{\omega}$ versus angle and $\frac{J}{\omega}$ for N=40.

The energy versus angle was plotted to find the magnetic easy and hard directions for each number of spin layers by keeping all the magnetic parameters fixed at $\frac{H_{in}}{\omega} = \frac{J}{\omega} = \frac{N_d}{\mu_0 \omega} = \frac{H_{out}}{\omega} = \frac{K_s}{\omega} = \frac{D_m^{(2)}}{\omega} = \frac{D_m^{(4)}}{\omega} 10$. The maxima and minima of the energy curve are related to magnetic hard and easy directions, respectively. The magnetic easy direction, hard directions and corresponding energy values are given in table 1. The easy axis gradually rotates from out of plane to in plane direction as the number of layer is increased. The energy required to rotate spins from magnetic easy direction to hard direction (or vice versa) gradually increases with the number of layers. It means that the magnetic anisotropy energy increases with the number of layers. Because the total number of spins in the film increases with the number of spin layers, the energy required to rotate all the spins also increases with the number of spin layers. Magnetic thin films with lower and higher magnetic anisotropies are useful in the applications of soft and hard magnetic materials, respectively.

| Number of spin layers | $\theta$ (easy) in degrees | $E/\omega$ (easy) | $\Delta E = E/\omega$(easy) - $E/\omega$(hard) | $\Delta\theta = \theta$(hard) - $\theta$(easy) in degrees |
|---|---|---|---|---|
| 10 | 12.5993 | -232 | 548.1 | 102.6225 |
| 20 | 26.9978 | -1050 | 726.2 | 109.8245 |
| 30 | 26.9978 | -1580 | 1088.9 | 107.9911 |
| 40 | 28.8026 | -2112 | 1452.7 | 106.1863 |
| 50 | 28.8026 | -2642 | 1815.7 | 106.1863 |

Table 1: Easy direction and magnetic energy for different number of spin layers.

Orientation of magnetic easy axis experimentally depends on deposition temperature, sputtering gas pressure, deposition rate, distance from target to substrate and number of layers [21].

Variation of magnetic easy axis orientation with deposition temperature was theoretically explained using Heisenberg Hamiltonian and spin reorientation [22, 23, 24]. According to the experimental data of sputtered Ni ferromagnetic films and electron beam evaporated Fe ferromagnetic films, the magnetic easy axis rotates from perpendicular to in plane direction as the thickness of the film is increased [25, 26]. This implies that our theoretical data presented in this manuscript agree with the experimental data of ferromagnetic thin films.

**4. Conclusion:**

One of the energy maximums and one of the energy minimums of 3-D plot of energy versus angle and stress induced anisotropy for N=40 were found at $\frac{K_s}{\omega}=16$ and $\frac{K_s}{\omega}=9$, respectively. By plotting graphs of energy versus angle at $\frac{K_s}{\omega}=9$ and $\frac{K_s}{\omega}=16$, magnetic easy and hard directions were found to be $10.8002^0$ and $120.6019^0$, respectively. Using the 3-D plot energy versus angle and spin exchange interaction for N=40, one energy minima and a maxima were observed at $\frac{J}{\omega}=4$ and $\frac{J}{\omega}=5$, respectively. Hence the magnetic easy and hard directions were found to be $12.5993^0$ and $115.2218^0$, respectively. The magnetic easy axis gradually rotates from $12.5993^0$ to $28.8026^0$, as the number of spin layers is increased from 10 to 50. This means that magnetic easy axis rotates from out of plane to in plane direction of the film. Magnetic anisotropy energy gradually increases from 548.1 to 1815.7, as the number of spin layers is increased from 10 to 50. However, the angle between magnetic easy and hard axis doesn't change considerably.